\documentclass[amsmath,amssymb,floatfix,12pt]{revtex4}
\newif\ifbw
\bwfalse

\usepackage{rotating}
\usepackage{graphicx}
\usepackage{verbatim}
\usepackage{amsfonts}
\usepackage{amsmath}
\usepackage{amssymb}
\usepackage{amscd}
\usepackage{ctable}

\newcommand \rt {\right}
\newcommand \lt {\left} 
\newcommand \nline {\nonumber \\}

\newcommand \freem {{\cal F}}

\newcommand \pxpy[2] {\frac{\partial #1}{\partial #2}}
\newcommand \dxdy[2] {\frac{d #1}{d #2}}

\newcommand \vxvy[2] {\frac{\delta #1}{\delta #2}}

\newcommand{\mbfr}{{\mathbf r}}

\begin{document}

\title{A Phase Field Crystal study of Solute Trapping}
\author{Harith Humadi}
\affiliation{Department of Materials Science and Engineering and
Brockhouse Institute for Materials Research, McMaster University,
1280 Main Street West, Hamilton, Canada L8S-4L7}

\author{Jeffrey J.~Hoyt}
\affiliation{Department of Materials Science and Engineering and
Brockhouse Institute for Materials Research, McMaster University,
1280 Main Street West, Hamilton, Canada L8S-4L7}

\author{Nikolas Provatas}
\affiliation{Department of Physics, Centre for the Physics of Materials, McGill University, 3600 Rue University, Montreal, Canada H3A 2T8}

\maketitle

\section{Abstract}
In this study we have incorporated two time scales into the phase field crystal model of a binary alloy to explore different solute trapping properties as a function of crystal-melt interface velocity. With only diffusive dynamics, we demonstrate that the segregation coefficient, $K$ as a function of velocity for a binary alloy is consistent with the model of Kaplan and Aziz where $K$ approaches unity in the limit of infinite velocity.  However, with the introduction of wave like dynamics in both the density and concentration fields, the trapping follows the kinetics proposed by Sobolev~\cite{sob95}, where complete trapping occurs at a finite velocity.
\section{Introduction}

Due to the relationship between complex dentritic microstructures and the mechanical properties of welded, soldered and cast components, solidification of alloys is one of the most researched subjects in materials science. Computational modeling of the solidification process has been an integral part of understanding the underlying physics that is associated with all aspects of solidification. In the past few decades, modeling of rapid solidification and solute trapping has gained attention for its role in many modern solidification processes such as thermal spraying, spin coating and laser melting~\cite{cahn69, introbook}. As the solid-liquid interface position advances at very fast rates ($\sim1$ $m/s$), there is substantial aberration from the equilibrium concentrations at the interface in both the solid and liquid phases. This deviation can be characterized through the segregation coefficient $K$, which is defined as the ratio of the solid concentration $C_{s}$ to the liquid concentration $C_{l}$ at the interface. The increase in the segregation coefficient with increasing interface velocity $V$ is known as solute trapping and a complete understanding of the trapping process requires the dependence $K(V)$ on the kinetic and thermodynamic properties of the alloy. 

There are three main theories of solute trapping. The most widely known is the continuous growth model (CGM) of Aziz and coworkers~\cite{aziz88,Azizbot93}. The CGM introduces the so-called diffusive speed, $V_{D}$, which is the velocity at which a solute atom can traverse the solid-liquid boundary. Aziz and Kaplan define this parameter as $V_{D}=D/\lambda$, where D is the diffusion coefficient in the interface region and $\lambda$ is the width of the interface. According to the CGM, significant trapping occurs when the interface velocity becomes comparable to $V_{D}$ and the main prediction of the model in the limit of dilute concentrations is given by: 
\begin{equation}
K(V)=\frac{k_{e}+V/V_{D}}{1+V/V_{D}}
\label{azizkv}
\end{equation}
where $k_{e}$ is the equilibrium segregation coefficient. Notice, that the complete trapping limit given by $K(V)=1$ is approached asymptotically as $V \rightarrow \infty$. The work of Jackson et al~\cite{jackson82}demonstrates similar asymptotic behavior, but $K(V)$ is described as a power law given by:
\begin{equation}
K(V)=k_{e}^{1/(1+AV)}
\end{equation}
Where $A$ is a parameter analogous to the reciprocal of $V_{D}$.

 In contrast, the work of Sobolev~\cite{sob95,sob97} predicts that there is an abrupt change in the segregation coefficient at a finite velocity. In other words, complete trapping ($K(V)=1$) occurs at a well defined velocity. Their theory can be explained in terms of a ``caging'' phenomenon in the bulk liquid of the system. A solute atom spends a short amount of time ($\sim$ pico seconds) in a cage formed by nearest-neighbor atoms before it undergoes a random hop to another cage-like shell. If the interface velocity $V$ is sufficiently fast, the liquid atoms attach themselves to the solid crystal at a rate that is comparable to this short time scale. At very high velocities the atoms do not have enough time to escape their cage and they completely freeze in their location. To capture the short time scale, Sobolev modified Fick's first law of diffusion by introducing the following:
 \begin{equation}
J+ \tau_{B} \pxpy{J}{t}=-D \nabla C
\label{sobflux}
\end{equation}
where $\tau_{B}$ is the relaxation time required for the flux to reach its steady-state regime. The fast atomic interactions in the liquid are incorporated into the time scale $ \tau_{B}$, which can be defined as $ \tau_{B}=\frac{D}{{V_{D}^{B}}^2}$ where $V_{D}^{B}$ is the bulk liquid diffusive speed.  $V_{D}^{B}$ describes how fast a solute atom can travel in the liquid. Sobolev emphasizes that solute atoms have a finite velocity in the liquid, which differs from the Aziz's model where the diffusion speed in the bulk liquid is assumed infinite. Combining Eq.~\ref{sobflux} with the conservation law $\pxpy{C}{t}=-\nabla \cdot J$, the diffusion equation in a steady state co-moving reference frame ($x=x_o-Vt$) can be written as:

\begin{equation}
 D \bigg( 1-\frac{V^2}{{V_{D}^{B}}^2}\bigg ) \pxpy{^2C}{x^2} + V \pxpy{C}{x} =0
 \label{movdiff}
\end{equation}

The solution of Eq.~\ref{movdiff} for the solute concentration ahead of the interface ($x > 0$) is:

\begin{align}
C(x)&=(C_{i}-C_{o}) \exp\bigg (-\frac{Vx}{D[1-(V/V_{D}^{B})^{2}]} \bigg ) + C_{o} & V < V_{D}^{B} \nline
C(x)&=C_{o} & V \ge V_{D}^{B}
\label{concsobsol}
\end{align}
When $x=0$ (interface position), the concentration is defined as $C_{i}$. As $x \rightarrow \infty$ (far bulk liquid position), the far field concentration is $C_{o}$. The solution given by Eq.~\ref{concsobsol} is similar to the well known solution of the concentration field in the liquid ahead of a moving interface except now the diffusion coefficient can be replaced by effective diffusion coefficient $D^{*}=D[1-(V/V_{D}^{B})^{2}]$. For the detailed solution to Eq.~\ref{movdiff}, the reader should consult the following references \cite{sob95, sob97}.

The above model extends the Aziz model Eq.~\ref{azizkv} by substituting the diffusion coefficient with the effective one, $D^{*}$, that was obtained from Eq.~\ref{concsobsol}. Therefore, the segregation coefficient in this case reaches unity when $ V\ge V_{D}^{B}$.  The Sobolev/Galenko model has the following formulation:

\begin{align}
K(V)&=\frac{k_{e}[1-(V/V_{D}^{B})^{2}]+V/V_{D}}{1-(V/V_{D}^{B})^{2}+V/V_{D}} &V<V_{D}^{B} \nline
K(V)&=1  &V\ge V_{D}^{B}
\label{kvsob}
\end{align}

As the velocity of the interface increases signifcantly, the atoms in the liquid are not able to escape from the solid back to the liquid and so they freeze in their positions as the system solidifies. In other word, the solute atoms do not have enough time to jump back to the liquid once the interface velocity equivalent to the bulk diffusive velocity, which results in complete trapping. 

There is mounting evidence that the Sobolev description of solute trapping is an accurate one. Danilov and Nestler~\cite{daninest} showed that the Sobolev prediction was a better fit for the experimental data on Si-As alloys by Kittl et al~\cite{kittl95}, especially in the high velocity limit. More recently, splat cooling experiments of Al-Mg were performed by Galenko and Herlach~\cite{galenkoherlach06} and the results show a change from a eutectic to super saturated solid solution at a finite velocity. Furthermore, theoretical work of~\cite{galenkodani} shows that the interface velocity as a function of undercooling changes from a power law to a linear relationship due to complete trapping and the model shows a very good agreement with Cu-Ni experiments. In addition to experimental studies, early Molecular Dynamics (MD) work by Cook and Clancy showed partitionless crystal growth ($K(V) \approx 1$) in a Leonard-Jones system at a finite velocity~\cite{cookclancy}. Recently, Yang et al also showed that complete trapping does occur at finite velocity. The authors simulated a Leonard-Jones binary and a Cu-Ni EAM model to show that the results hold for different systems~\cite{yangprl}. It was found that the Sobolev model was a better fit to their MD data, while the Aziz model underestimated the high velocity cases when $K(V)=1$. The Yang et al. results confirmed complete trapping but MD simulations are atomistic and cannot show whether there is a need for the second time derivative in the continuum description that was suggested by Sobolev~\cite{sob95}. 

The phase field models of Wheeler et al~\cite{BMW} and Echebarria el al~\cite{EFKP} do not show complete trapping in the high velocity limit using the parabolic form of the dynamic equations. Galenko et al modified these models to their hyperbolic form i.e. adding the second time derivative to the concentration field~\cite{galenkosolute11}. They observed that $K(V)$ does indeed tend to $1$, however its behavior varies depending on the phase field phenomenology they choose. In addition, the diffusion coefficient of these models is a function that can be tailored to produce different trapping behaviors. 

In this paper we study solute trapping using a phase field crystal (PFC) alloy model with inertial dynamics. The PFC model captures the atomic scale structure analogous to MD yet its diffusive time scale is of the same order as regular phase field models. Also, the PFC is motivated from classical density functional theory and contains only a minimal number of parameters, which in principle, can be derived from fundamental liquid state properties through the direct correlation functions controlling the excess energy. These features are relevant to solute trapping since most of the interactions occur on the liquid side of the solid-liquid interface. Moreover, changing the parameters and examining their effects on the system is rather fast compared to MD. For instance, the mobility of the system can be adjusted without the need for new interatomic potential(s) in order to see the effect of the change on the trapping behavior. The work of Stefanovic et al~\cite{peterprl,peterpre} was the first to add a second time derivative to the pure PFC model~\cite{elder2002}. They motivated the second time derivative from hydrodynamics and used the resulting kinetic equations to study the deformation and plasticity in nano-crystals. The addition of the second time derivative to a two component alloy PFC model will be discussed in the model section of this paper. The competition between inertial dynamics and mobility in both the concentration and density field dictates the trapping behavior of the solute atoms will be examined. 

The rest of the paper is organized as follows. Section~\ref{model} describes the PFC model free energy as well and the equations of motion for the density and concentration. Section \ref{results} presents our findings by examining the relationship between the wave terms, the mobility of the system and their effect on the concentration and density fields. We end the paper with conclusions in Section \ref{conclusions}.


 \section{Modeling Approach}
 \label{model}

 \subsection{PFC Model}
For a binary alloy system that consists of A and B atoms we start by introducing the PFC free energy functional~\cite{elder2007}:
\begin{align}
\frac{\freem}{k_BT\rho_{\ell} R^d}\! \!=\!\! \int
d \mbfr \bigg\{B^{\ell}\frac{n^2}{2}+B^x\frac{n}{2}\lt(2\nabla^2+\nabla^4\rt)n-\frac{t}{3}n^3+\frac{\nu}{4}n^4+\frac{\omega}{2}\psi^2+\frac{u}{4}\psi^4+\frac{C}{2}|\vec{\nabla}\psi|^2\bigg\}
\label{eldPFCalEnrgy}
\end{align}
The fields $n$ and $\psi$  are the dimensionless local number density and concentration, respectively and they can be approximated as follows:
\begin{align}
n&=\frac{\rho_{A}+\rho_{B}}{\rho_{\ell}}-1&&\psi=\frac{\rho_{A}-\rho_{B}}{\bar{\rho}}
\label{ncdefs}
\end{align}
$\rho_{A}$ is the number density of species A (Solvent) and $\rho_{B}$ is the density of species B (Solute),  $\bar{\rho}$ is the average density, $\rho_{\ell}$ is a reference liquid density, $k_B$ is Boltzmann's constant, $R$ is the average atomic radius, which sets the length scale of the system, and $T$ is the temperature. The last three terms in Eq.~\ref{eldPFCalEnrgy} represent a free energy of the Cahn-Hilliard type~\cite{cahn58}, which has been used extensively to study phase separation. $B^{\ell}$ is the dimensionless bulk modulus of the liquid and $B^x$ is the dimensionless bulk modulus of the solid, which control the energy scale of the system. Following~\cite{elder2007} they are expanded as follows $B_\ell=B_0^\ell+B_2^\ell\psi^2$ and $B^x=B_0^x$. The difference between the bulk moduli $\Delta B_0 = B_0^\ell - B_0^x$ sets the temperature scale of the system. $\nu$,$t$ and $u$ are constant parameters related to the direct two point correlation functions and they can be calculated using classical density functional theory or can be fitted to phenomenological models to describe various materials properties. It is worth noting that minimization of the free energy yields hexagonal structures in 2D, BCC in 3D and a constant density profile in the liquid. The model is capable of producing two different phase diagrams, eutectic and spinodal binary systems. Eq.~\ref{eldPFCalEnrgy} is a reasonably simple model that can be used to simulate solidification, phase segregation and elasticity/plasticity~\cite{niksbook}. 

Diffusion-controlled solidification dynamics is the PFC formalism are simulated via:
\begin{align}
\pxpy{\rho_{A}}{t}&=\vec{\nabla} \cdot \bigg(M_{A} \vec{\nabla}\vxvy{\mathcal{F}}{\rho_{A}}\bigg) \nline
\pxpy{\rho_{B}}{t}&=\vec{\nabla} \cdot \bigg(M_{B} \vec{\nabla}\vxvy{\mathcal{F}}{\rho_{B}}\bigg)
\label{abdynamics}
\end{align}
where $M_{A}$ and $M_{B}$ are the mobilities of each atomic species. The dynamics is driven by minimizing the free energy of the system in similar fashion to the Cahn-Hilliard equation~\cite{cahn58}. One can rewrite Eqs.~\ref{abdynamics} in terms of the density $n$ and the concentration $\psi$ fields using Eqs.~\ref{ncdefs}~\cite{elder2007}:

\begin{align}
\pxpy{n}{t}&=\vec{\nabla} \cdot M_{1} \vec{\nabla}\vxvy{\mathcal{F}}{n} + \vec{\nabla} \cdot M_{2} \vec{\nabla}\vxvy{\mathcal{F}}{\psi} \nline
\pxpy{\psi}{t}&=\vec{\nabla} \cdot M_{1} \vec{\nabla}\vxvy{\mathcal{F}}{\psi} + \vec{\nabla} \cdot M_{2} \vec{\nabla}\vxvy{\mathcal{F}}{n}
\label{ncdynamics}
\end{align}
where $M_{1}=(M_{A}+M_{B})/ \rho_{\ell}^{2}$ and  $M_{2}=(M_{A}-M_{B})/ \rho_{\ell}^{2}$. Notice, we will neglect the effects of thermal fluctuations in all of the simulations. For substitutional diffusion we assume the mobilities of species A and B are the same, which implies that $M_{2}=0$ and the cross terms vanish from Eqs.~\ref{ncdynamics}. 

As mentioned earlier, Sobolev~\cite{sob95} introduces a second time derivative in concentration to capture the fast interactions in the bulk liquid. In the case of PFC alloys, Stefanovic et. al considered inertial dynamics in both the concentration and density field ~\cite{peterthesis}. They derived these as an extension of this concept in pure materials \cite{peterprl} to binary alloys. Equations of motion for the concentration and density in this limit are given by:
 \begin{align}
\beta\pxpy{^2 n}{t^2}+\pxpy{n}{t}&=M\nabla^2\bigg(\vxvy{\mathcal{F}}{n}\bigg) \nline
\gamma\pxpy{^2 \psi}{t^2}+\pxpy{\psi}{t}&=M\nabla^2\bigg(\vxvy{\mathcal{F}}{\psi}\bigg)
\label{wavedynamics}
\end{align}
where $M=M_1$, the $\beta$ and $\gamma$ coefficients are responsible for introducing two time scales in the dynamics of $n$ and $\psi$. The ratio of $\frac{\beta}{M}$ and $\frac{\gamma}{M}$ will be shown to be important quantities that will be discussed in depth in the next section. Each ratio controls the effect of atomic inertial interactions in its respective field. 

\subsection{Numerical Methodology}

The simulation domain is a $2D$ channel consisting of 4096 grid points in the ($x$) direction normal to the solid-liquid interface and 64 grid points in the parallel direction ($y$). To initiate the simulation, a solid seed (periodic density) of $(32x \times 64y$  was positioned at the far left part of the simulation box while the rest of the simulation box was filled with liquid (constant density). All simulations were performed with periodic boundary conditions in both the $x$ and the $y$ directions. The average concentration was set to be $\psi_0=-0.175$ and the temperature $\Delta B_{0}=B_0^\ell-B_0^x=0.08$. As mentioned in the introduction, the Aziz model (Eq.~\ref{azizkv}) and the Sobolev (Eq.~\ref{kvsob}) model are derived for the dilute limit. The parameters above are thus chosen to ensure the system has a dilute concentration of $B$ atoms and the average concentration is such that it lies in the solid-liquid coexistence region. 

\begin{figure}[htbp]
\resizebox{5in}{!}{\includegraphics{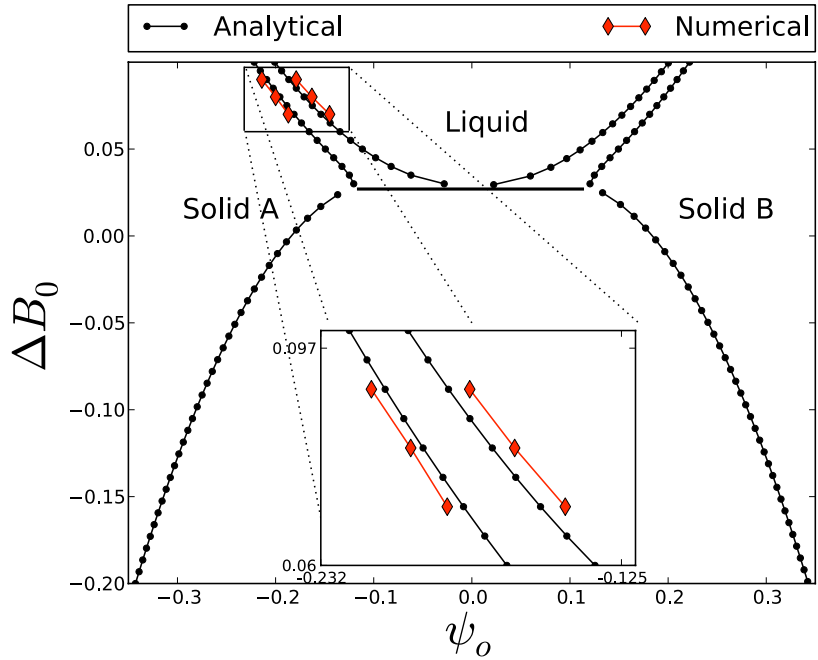}}
\caption{Eutectic phase diagram plot of $\Delta B_0$ vs  $\psi_o$ for the parameters $B_0^x=1.00$, $B_{2}^{\ell}=-1.80$, $t=0.60$, $\nu=1.00$, $u=4.00$ and $\omega=0.008$. The insert shows the difference between the analytical and simulated coexistence lines of the phase diagram.}
\label{fphasediagram}
\end{figure}

Simulations were done for a eutectic phase diagram shown in Fig~\ref{fphasediagram}. Since the model only supports hexagonal symmetry in $2D$ and the simulation box is of a rectangular shape, some stress is present in the system, which can cause a change in the equilibrium concentration. To account for this, and any numerical inaccuracies associated with analytically calculating the phase diagram, we numerically evaluated the coexistence lines that were used to conduct our simulations.  Analytical and numerical coexistence lines of the PFC alloy model are shown in Fig.~(\ref{fphasediagram}).
In particular,  the black lines represent the analytical solution obtained from an amplitude expansion technique as described in~\cite{elder2007} while the red lines show the the equilibrium solid/liquid concentrations obtained from the simulations. The equilibrium simulation was conducted using the simulation setup mentioned in the previous paragraph. Increasing the system size in the $y$-direction relieves most of the stress in the system but a plateau is reached after roughly $64$ grid points, with the simulation time increasing significantly for larger systems without a considerable amount of further stress relief. The remaining discrepancy between the analytic solution and the full numerical simulation shown in Fig~\ref{fphasediagram} is most likely due to the approximations involved in the amplitude expansion technique. It is noted that the concentration on the left hand side of the phase diagram has a negative value, which implies the absolute value of the solidus concentration to be larger than that of the liquidus. In the PFC definition of the concentration (Eq.~\ref{ncdefs}) a value of $\psi=0$ corresponds to an actual concentration of $0.5$. Therefore in the results presented below, we will add $1$ to both the solid and liquid interface compositions when reporting the segregation coefficient i.e. $K(V)=(\psi_s+1)/(\psi_l+1)$. 

\begin{figure}[htbp]
\resizebox{5in}{!}{\includegraphics{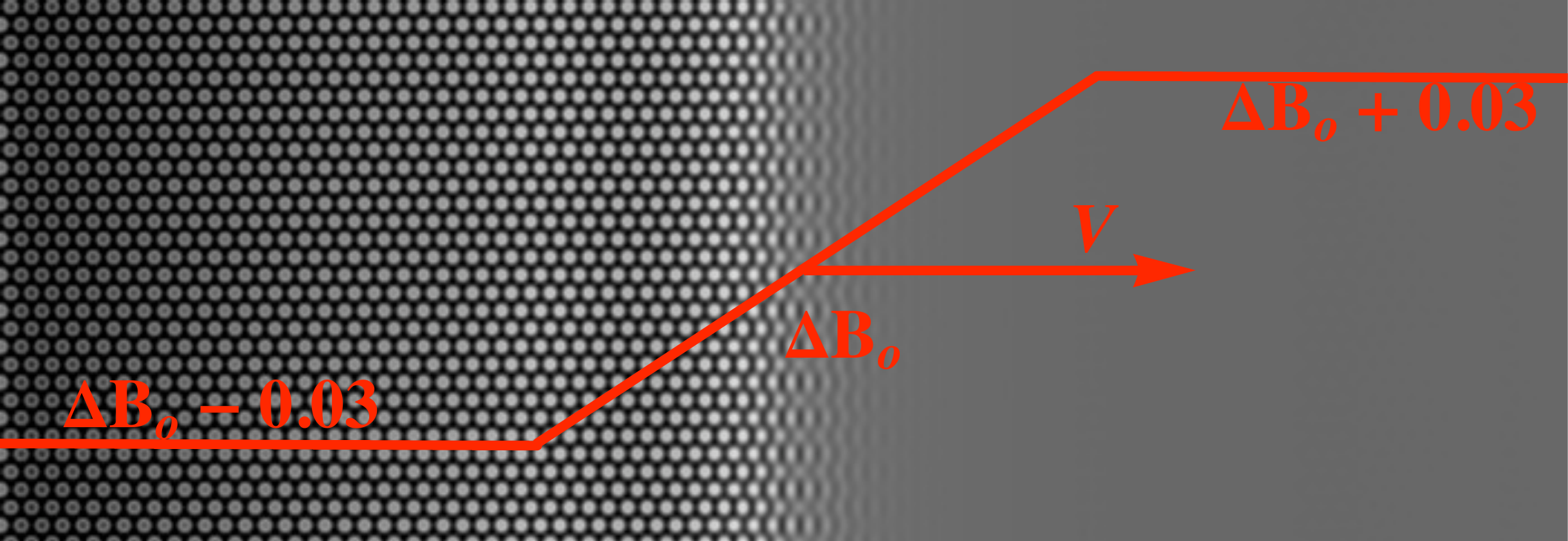}}
\caption{A typical 2D density profile with a temperature gradient and cap overlaid on top to illustrate the deriving force mechanism employed in the simulations.}
\label{gradtemp}
\end{figure}

The method of driving the solid-liquid interface employed in the current study involves the application of a linear temperature ($\Delta B_0$) gradient in the direction normal to the boundary. The temperature gradient (G) is chosen such that the liquid phase remains above the equilibrium liquidus temperature and the solid is maintained at temperatures below the liquidus. Over the course of the simulation, the gradient is translated in the $x$-direction at the specified velocity. Some care must be taken in applying the temperature gradient. The temperature gradient choice affects the solidification transient regime, either prolonging or shortening it, depending on the steepness of temperature gradient. Moreover, a temperature cap, that is, the horizontal portions of the temperature profile shown in Fig~\ref{gradtemp}, in the bulk must be incorporated. Certain temperature caps can be used to ensure that the temperature gradient does not ``outrun'' the solid-liquid interface. If the cap is too small, the temperature of the solid is too close to the starting temperature, which gives the interface limited driving force. Therefore, the temperature gradient will outrun the interface at large velocities, which causes the system to solidify isothermally. In order to pull the interface at very high velocities the interface and the solid portion of the system must be assigned lower temperature caps. Additionally, the temperature cap prevents entering the AB part of the phase diagram, which causes the formation of $\alpha$ and $\beta$ phases (i.e. precipitates). For the purpose of this paper the temperature cap and the temperature gradient were chosen to be $\Delta B_{o}=\pm0.03$ and $G=0.0002$ respectively.   

\section{Results and Discussion} 
\label{results}

 \subsection{Inertial Dynamics in the Concentration Field}
We start the discussion of solute trapping by examining the steady-state concentration profiles. To produce a 1D concentration profile, the concentrations are averaged in the $y$-direction parallel to the interface. As will be seen in the results presented below, the solute concentration profiles exhibit periodic oscillations in the solid phase on a scale of its inter-planar spacing. The periodicity in the solid part of the concentration profile is a result of the density-concentration coupling term in Eq.~\ref{eldPFCalEnrgy}. The average solid concentration  in the steady-state regime can be calculated by averaging over the atomic planes of the solidified crystal. Refer to the appendix~\ref{appa} for further explanation on the concentration profile calculations. 

\begin{figure}[htbp]

 \centering

  \begin{tabular}{cc}


    \includegraphics[width=80mm]{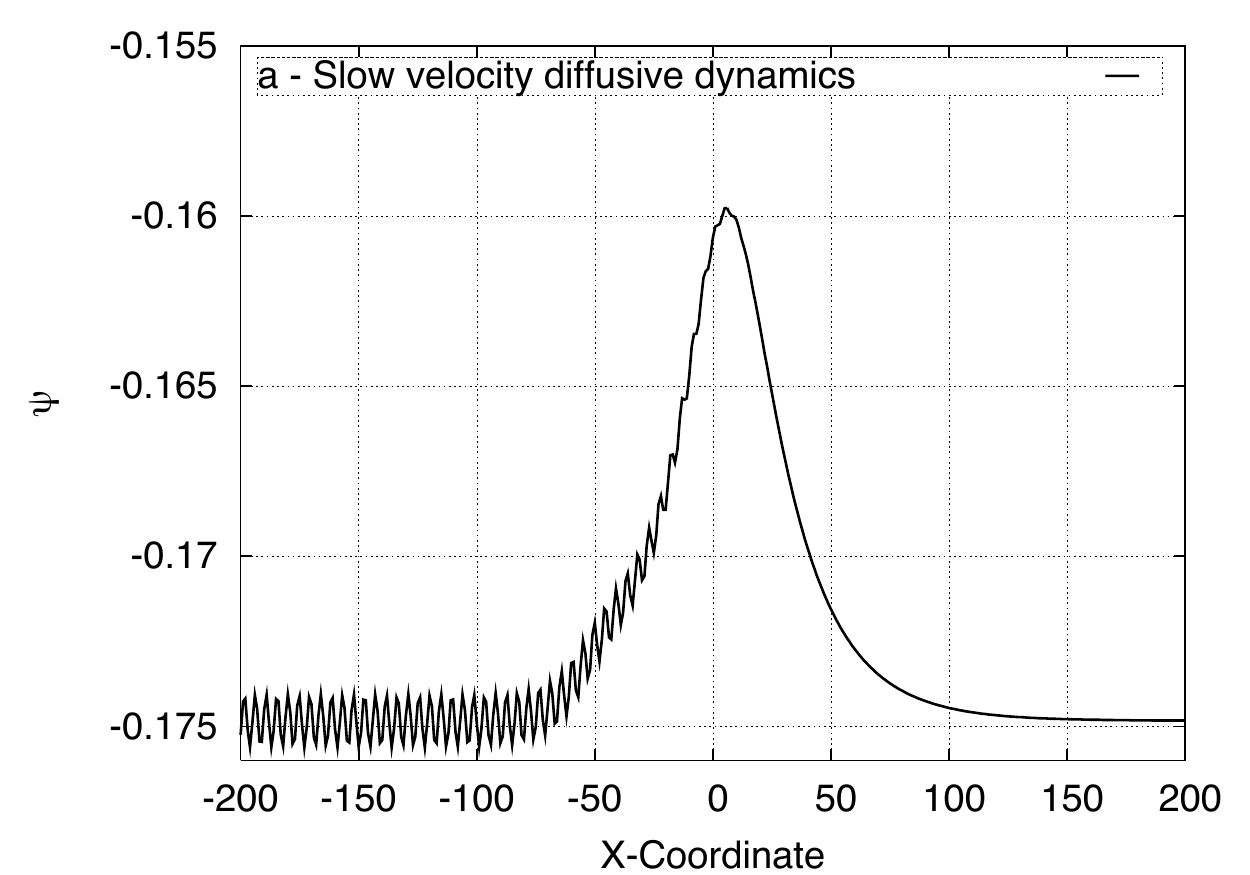}&

    \includegraphics[width=80mm]{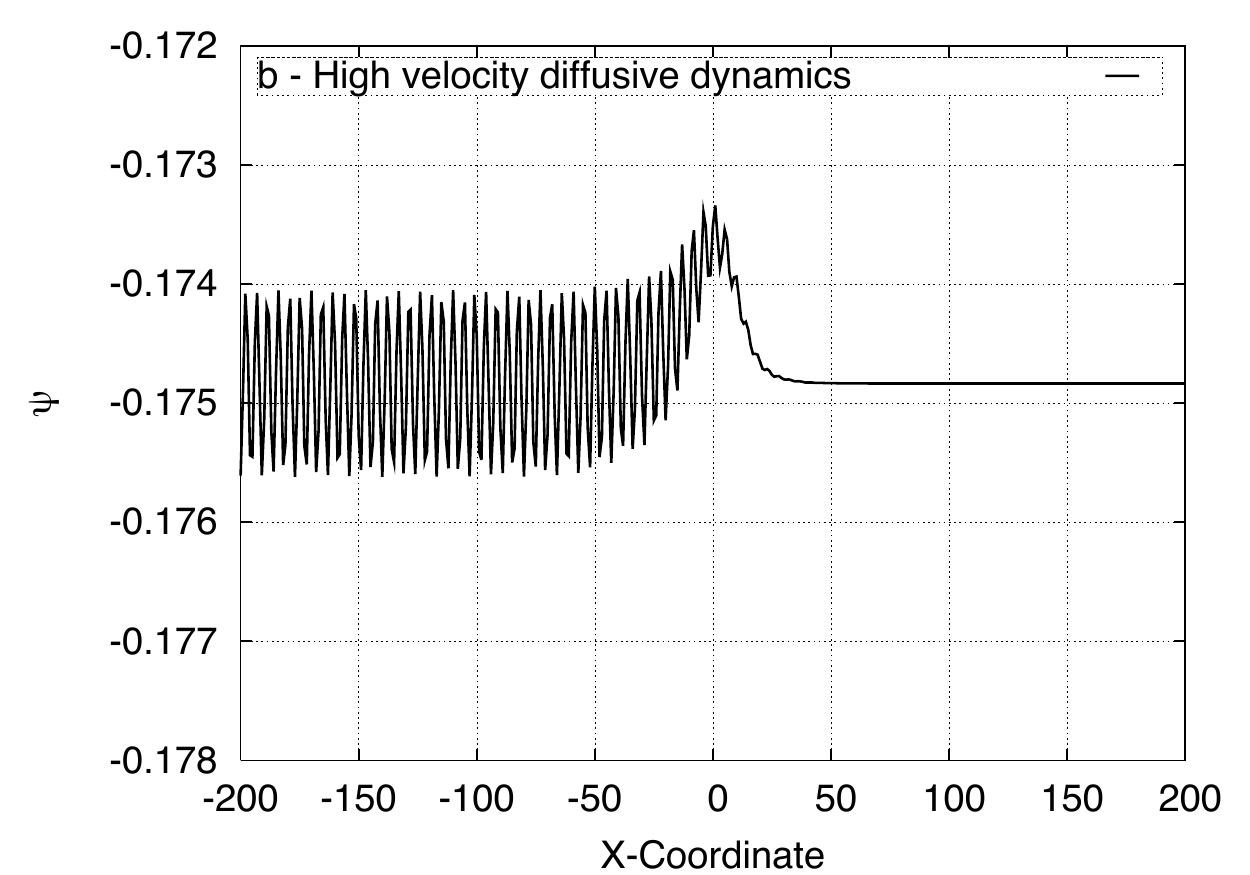}\\
    
     \includegraphics[width=80mm]{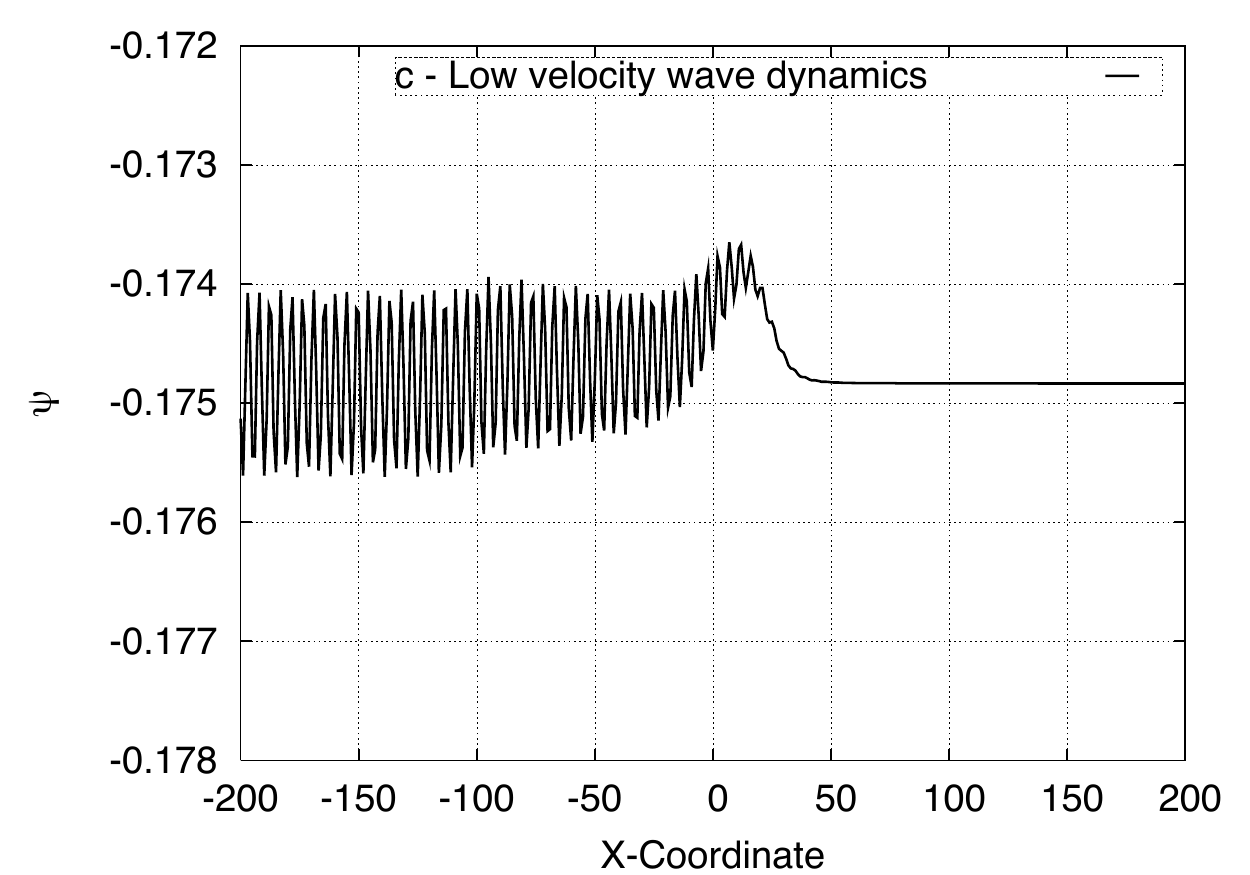}&
     
     \includegraphics[width=80mm]{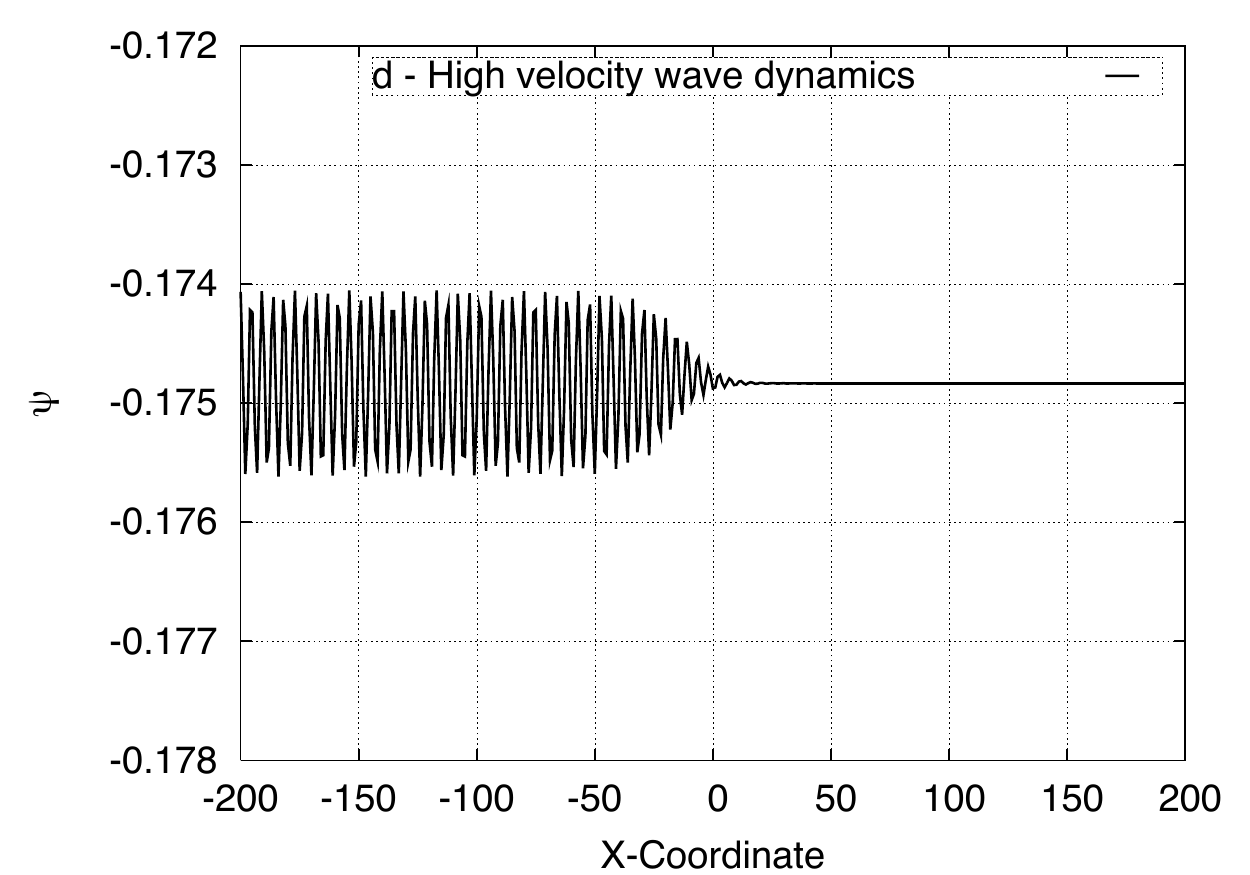}\\
     
      \includegraphics[width=80mm]{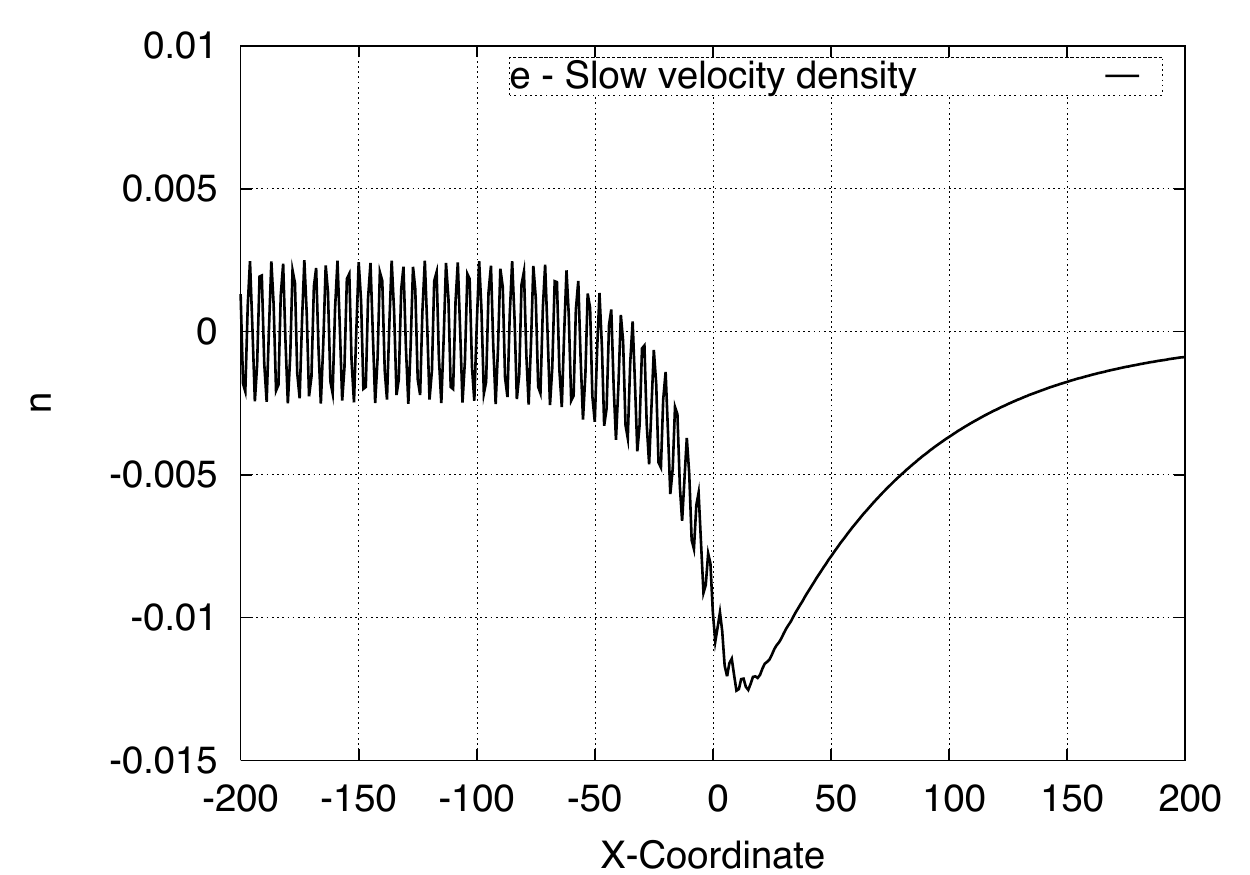}&
     
     \includegraphics[width=80mm]{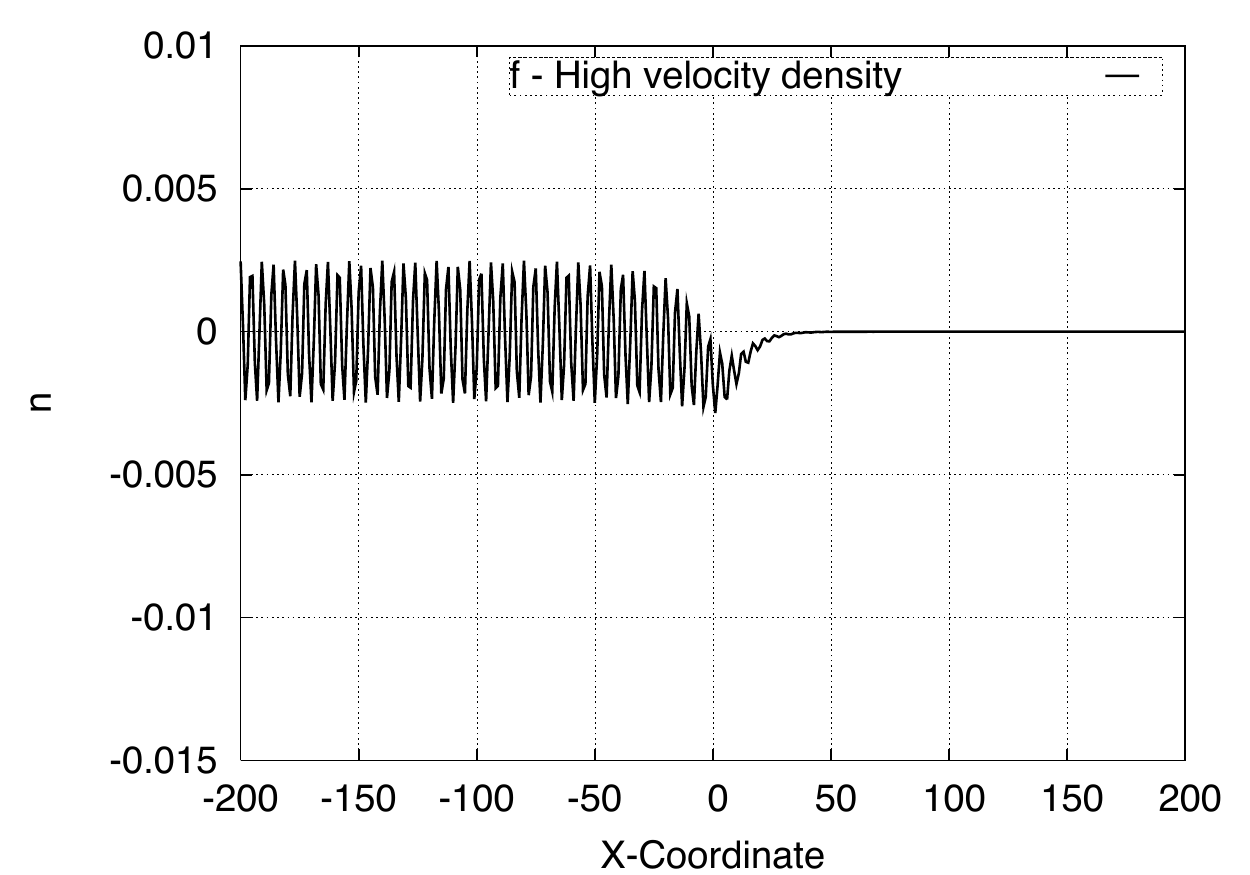}

 \end{tabular}

  \caption{Typical concentration profiles; a and b simulated with diffusive dynamics, while c and d simulated with inertial dynamics. Frames e and f illustrate the corresponding density profiles at (e) $V=0.2$ and (f) $V=0.5$, both simulated with diffusive dynamics.}
   \label{concprofiles}

\end{figure}

Fig~\ref{concprofiles}a and \ref{concprofiles}b show typical concentration profiles for slow ($V=0.02$) and fast ($V=0.5$) velocities, respectively, using diffusive kinetics($\beta=0, \gamma=0$). The solute accumulation ahead of the interface decreases as the velocity increases. It is clear that even in extremely fast velocities (Fig~\ref{concprofiles}b) the concentration ahead of the interface is higher than the solid concentration. When the second time derivative is activated in the concentration field ($\gamma/M=100.0$)  and $\beta=10$, there is a substantial amount of trapping in the low velocity profile as illustrated in Fig~\ref{concprofiles}c. One can define the occurrence of complete solute trapping when the liquid concentration ahead of the interface equals the averaged solid concentration through the interface. Fig~\ref{concprofiles}d displays a concentration profile where complete trapping occurs, it is evident the solute peak ahead of the interface vanishes. 

A summary of the PFC model parameters is given by Table~\ref{betagammam} for all the simulations reported in this study. The values of $\beta$ and $\gamma$ were chosen so the ratios of these quantities give insight in the solute trapping behavior. One set of simulations has diffusive dynamics in both fields to provide a baseline for the effect of the inertial dynamics on the trapping behavior. 

\begin{table}
  \centering\begin{tabular}{c*{6}{l}}
   \hline
    $\beta$             &$\gamma$     &M                  &$\frac{\beta}{M}$              & $\frac{\gamma}{M}$               &$\frac{\gamma}{\beta}$   &Comments\\
     \hline \hline

        0.00		 & 0.00 	              & 1.00	   		   & 0.00    		 & 0.00 	                & -- 		&Diffusive dynamics in both fields \\
        10.00          & 110.0                & 11.0      		   & 0.91			 & 10.0       	         & 11.0			& Wave dynamics in both fields\\
         14.0          & 2000.0    	      & 20.0		            & 0.71 			 & 100.0		          & 142.9 			& Wave dynamics in both fields\\			
        10.0          & 2000.0    	      & 20.0		            & 0.50 			 & 100.0		          & 200.0 			& Wave dynamics in both fields\\
         14.0          & 200.0    	      & 20.0		            & 0.71 			 & 10.0		          & 14.29 			& Wave dynamics in both fields\\			
        10.0          & 200.0    	      & 20.0		            & 0.50 			 & 10.0		          & 20.00			& Wave dynamics in both fields\\
        10.0          & 100.0 		      & 20.0		            & 0.50			 & 5.0		          & 10.00			& Wave dynamics in both fields\\
        10.0          & $66.\bar{66}$     & 20.0		            & 0.50			 & 3.$\bar{33}$	          & 6.$\bar{66}$        & Wave dynamics in both fields\\
        10.0          & $33.\bar{33}$     & 20.0	                    & 0.50			 & 1.$\bar{66}$	          & 3.$\bar{33}$			& Wave dynamics in both fields\\
        10.0          & $16.\bar{66}$     & 20.0   		   & 0.50			 & 0.8$\bar{3}$		  & 1.$\bar{66}$			& Wave dynamics in both fields\\
        \hline
   \end{tabular}
   \caption{Summary of simulation parameters in Eqs.~\ref{wavedynamics}}
   \label{betagammam}
\end{table}

\begin{figure}[htbp]
 \centering

  \begin{tabular}{cc}


    \includegraphics[width=80mm]{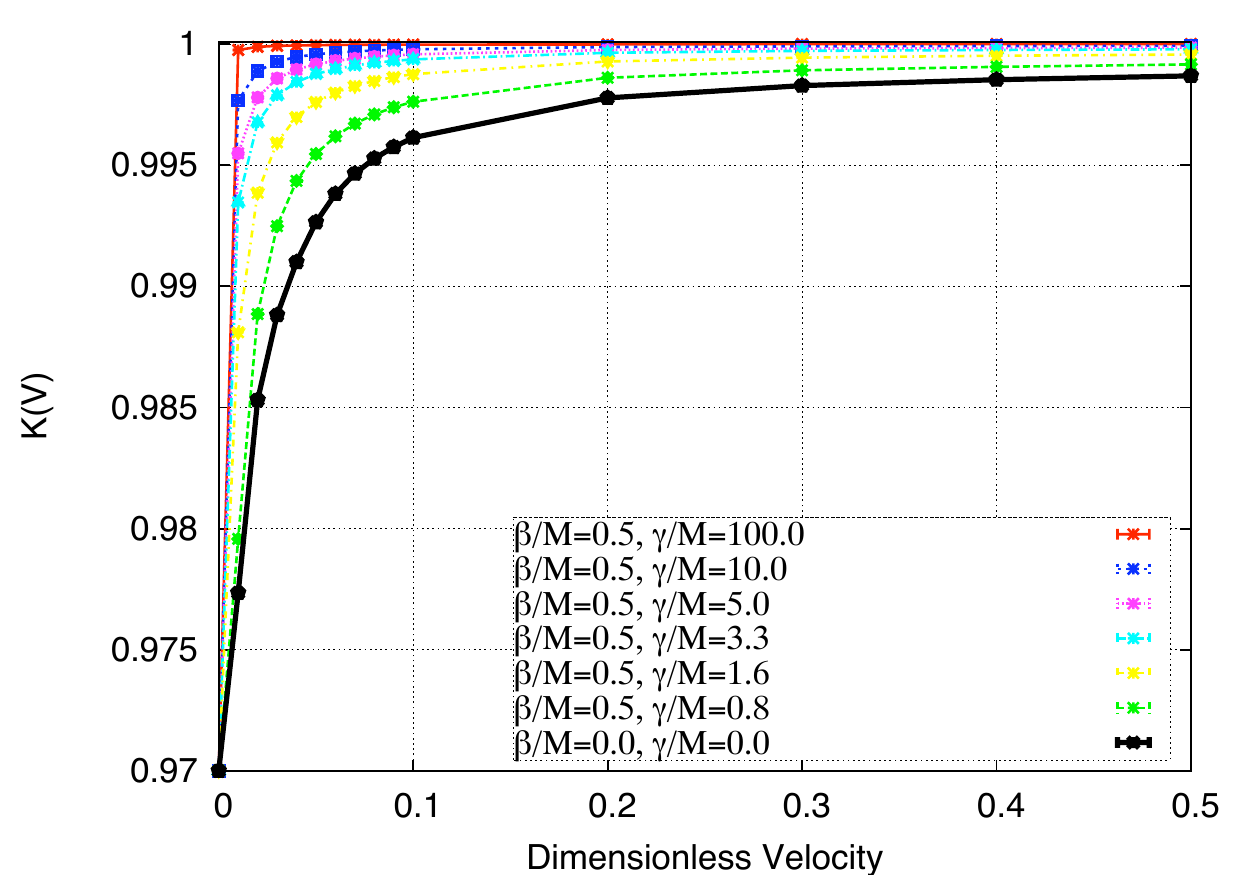}&

    \includegraphics[width=80mm]{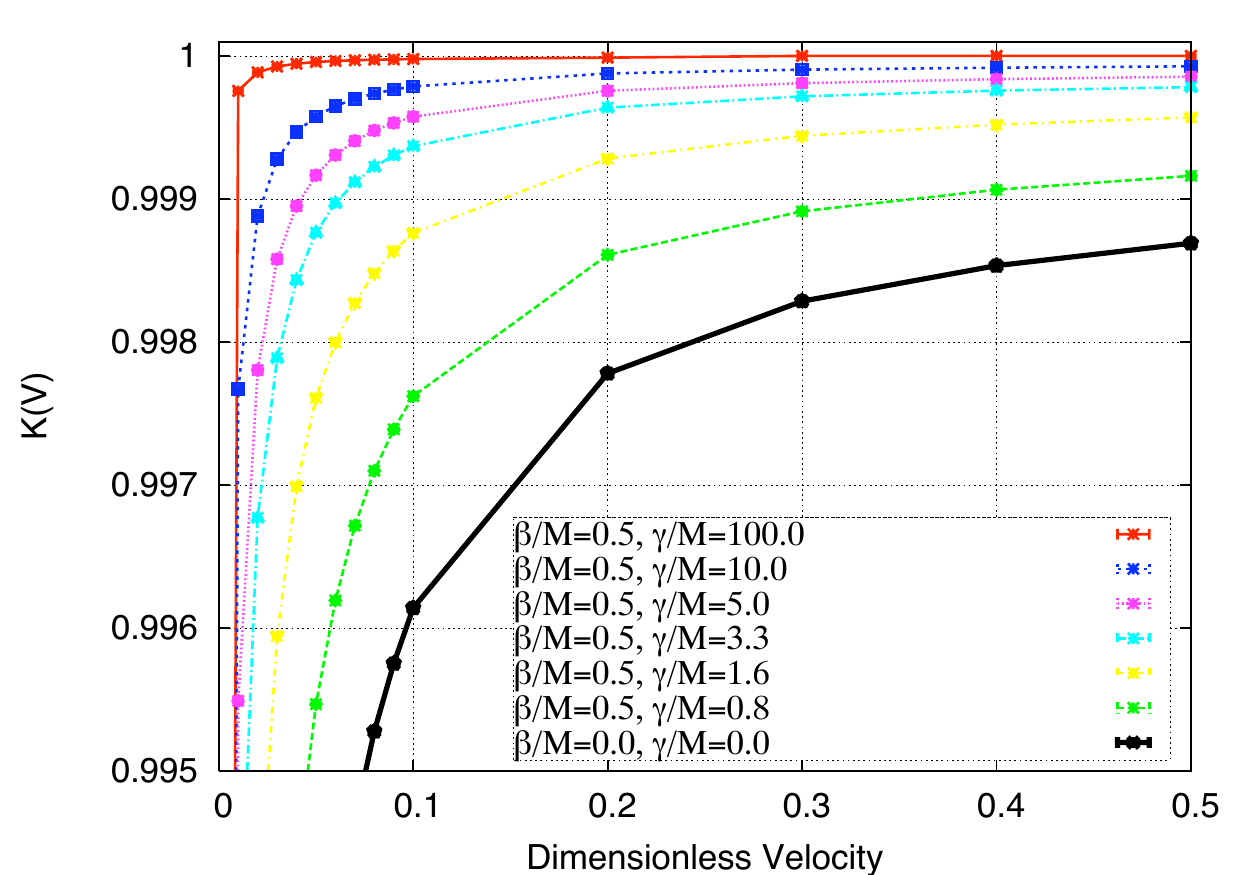}
    
 \end{tabular}
  
  \caption{K(V) vs. Velocity with constant $\beta$ and variable $\gamma$. The case of purely diffusive dynamics is also shown for comparison. The right panel is zoom in of the left panel. }
  \label{bkv}

\end{figure}

\begin{figure}[htbp]

 \centering

  \begin{tabular}{cc}


    \includegraphics[width=80mm]{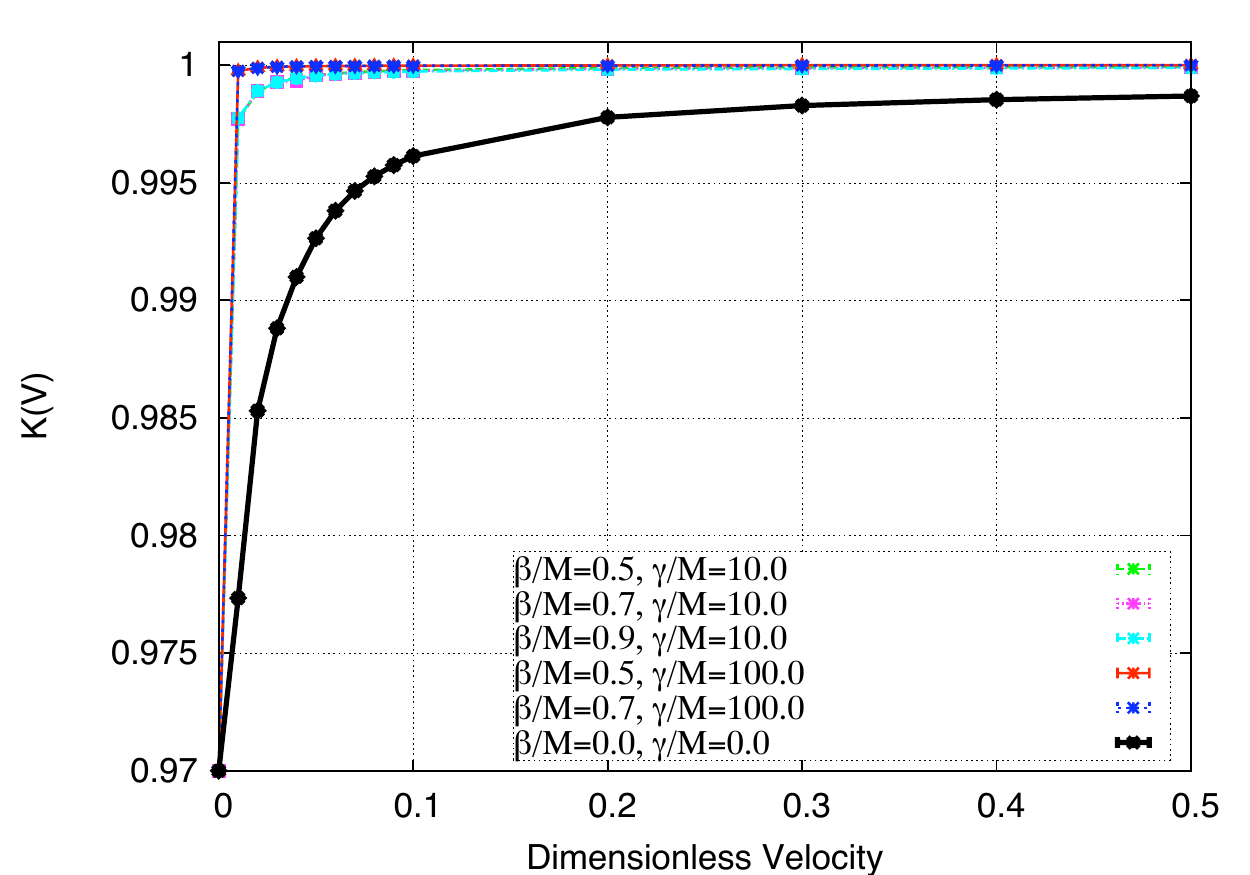}&

    \includegraphics[width=80mm]{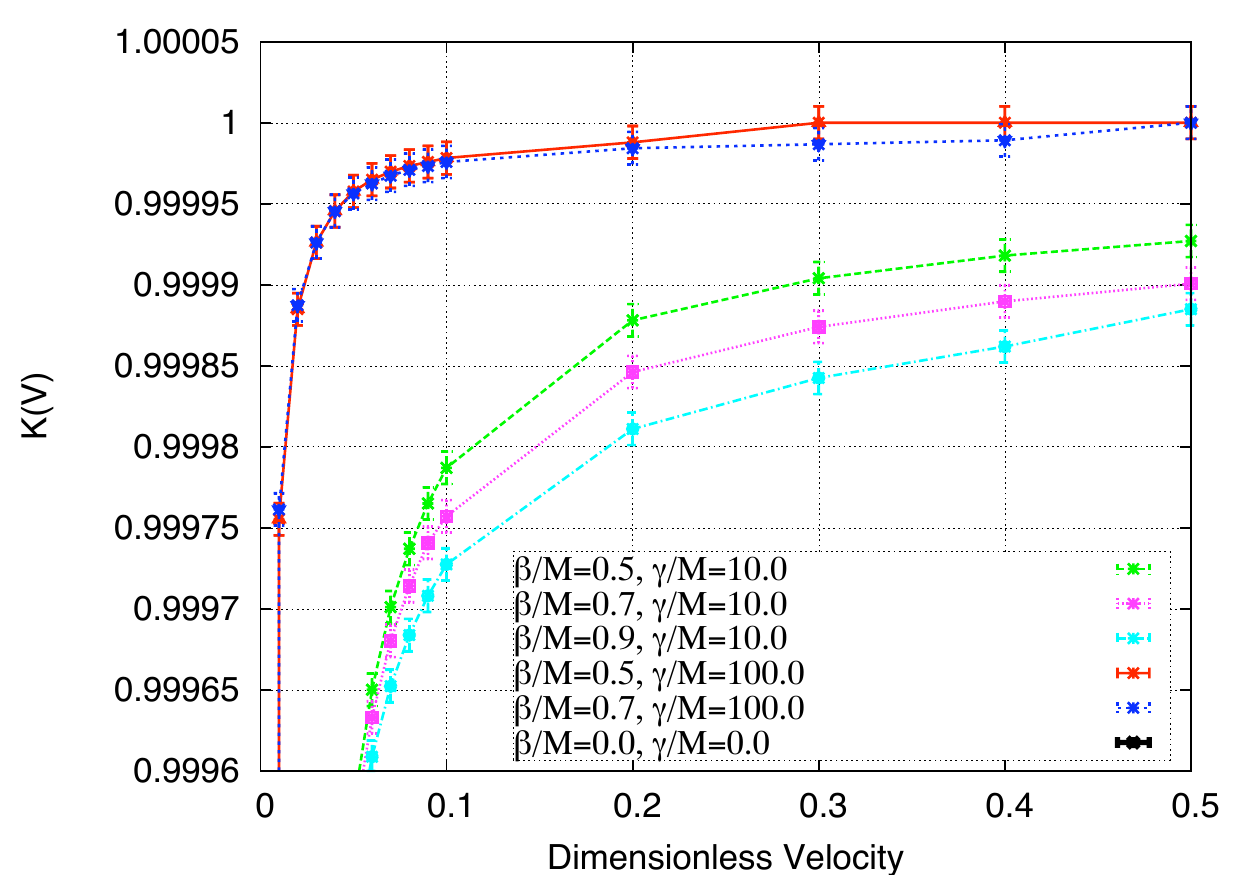}
    
 \end{tabular}
  
  \caption{K(V) vs. Velocity with two values of $\gamma$ and variable $\beta$. The case of purely diffusive dynamics is also shown for comparison. The right panel is zoom in of the left panel.}
  \label{gkv}

\end{figure}

As the velocity of the interface increases using diffusive dynamics in the concentration and density, $K(V)$ approaches unity in an asymptotic manner analogous to the Aziz model ($\gamma=0$, $\beta=0$ (black in color data )) in Fig~\ref{bkv} or Fig~\ref{gkv}. For the  same set of parameters, but the second time derivative is activated in both fields,  complete solute trapping can be achieved for a steady state front propagating at a finite velocity, as seen by the $\gamma/M=100$ and $\beta/M=0.5$ data (red in colour) in  Figs.~\ref{bkv} or ~\ref{gkv}. 
We also found analogous behaviour, leading to complete trapping, was exhibited with $\beta=0$.  Figure~\ref{bkv} also shows that, for fixed 
$\frac{\beta}{M}$, the amount of trapping rapidly increases with increasing $\frac{\gamma}{M}$. These observations  suggest that complete solute trapping can not take place in the alloy PFC model without the second time derivative being present in the concentration field.  The term $\gamma$ in the PFC context thus plays the same role as the 
$\tau_B$ parameter in Gelenko's theory. 

Conversely, it was  found that for a fixed value of  $\frac{\gamma}{M}$, increasing $\frac{\beta}{M}$ leads a reduction of trapping in the $\psi$-field over the same range of velocities as that discussed above.  This is illustrated in Fig~\ref{gkv}. The $\gamma/M=100$ data of Fig.~\ref{gkv} also suggest that an increase in $\beta/M$  requires a higher velocity to achieve complete trapping in the $\psi$ field. Lowering $\gamma$ to $10$ has the effect of pushing to even larger values the trapping velocity (i.e. increasing $V_D^B$).  Thus, for lower values of $\gamma$, we expect that the velocity needed to achieve complete trapping (as $\beta$  increases) increases for lower values of $\gamma$.  To better understand the role of the density field on the solute partitioning, it is instructive to consider the role that the inertial term of the density field plays in the in the kinetics of the interface. This will be examined in the next section.

  \subsection{Inertial Dynamics in the Density Field}
The behavior of diffusive density dynamics in rapid solidification was examined by~\cite{granasyprl} using the pure PFC model of Ref.~\cite{elder2002}. They show that as the velocity of the interface increases, the difference between the solid and liquid densities decreases. They demonstrate that the density field follows an Aziz type trapping, analogous to that of the concentration. They attribute this change in the density to vacancy trapping as the interface advances in rapid solidification. We observe similar behavior in the density field of the binary alloy, where in the slow velocity regime, there is a noticeable jump in the density across the the solid-liquid interface (Fig~\ref{concprofiles}e). As the the interface deviates from the local equilibrium condition at higher interface velocity, the density jump significantly drops, as shown in Fig~\ref{concprofiles}f. 

The use of inertial dynamics in the pure material to examine the transition from a periodic to homogenous solution was first discussed by Galenko et al~\cite{galenkopure}. In their work, they investigate the stability of the second order differential equation for both the Swift-Hohenberg and the PFC models. At low driving force (small interface speeds), the inertial terms can be neglected to retain a parabolic form of the dynamics, since the transition from one phase to another occurs on diffusive time scales. As the driving force increases, the inertial term in the dynamics (hyperbolic form) provides an extra degree of freedom to capture rapid kinetics. They compare the parabolic and hyperbolic solution for the same [high] driving regime and find the hyperbolic form of the equation leads to slower transition speeds. They ascribe the slowing down in the front speed to the increase in the relaxation time caused by the inertial kinetics. As the $\beta$ term increases in Eq.~\ref{wavedynamics}, the system takes a longer time to switch to diffusive dynamics.  We also simulated isothermal quench simulations (i.e. no temperature gradient) and similarly found that the solid-liquid interface slows down as the effect of the inertial term increases, and when $\beta/M$ is of order $100$ or higher, it was found that the interface virtually stops over the time of the simulation, leading to subsequent concentration pile up ahead of the interface,  and lower solute trapping. 

The situation is different for the case of directional solidification used here for binary alloys, since conditions at the solid-liquid interface are driven by a temperature gradient that imposes the front velocity at steady state. In this case, for a given set of parameters in the free energy, as the pulling velocity increases the interface temperature is found to decrease i.e. slides down the temperature gradient towards its lower cap (see Fig.~\ref{gradtemp}), effectively increasing the driving force to keep the interface moving with the imposed higher speed. It was also found that a decrease in local interface temperature occurred for a fixed interface velocity when $\beta/M$ was increased. 
\begin{figure}[htbp]
\includegraphics[width=100mm]{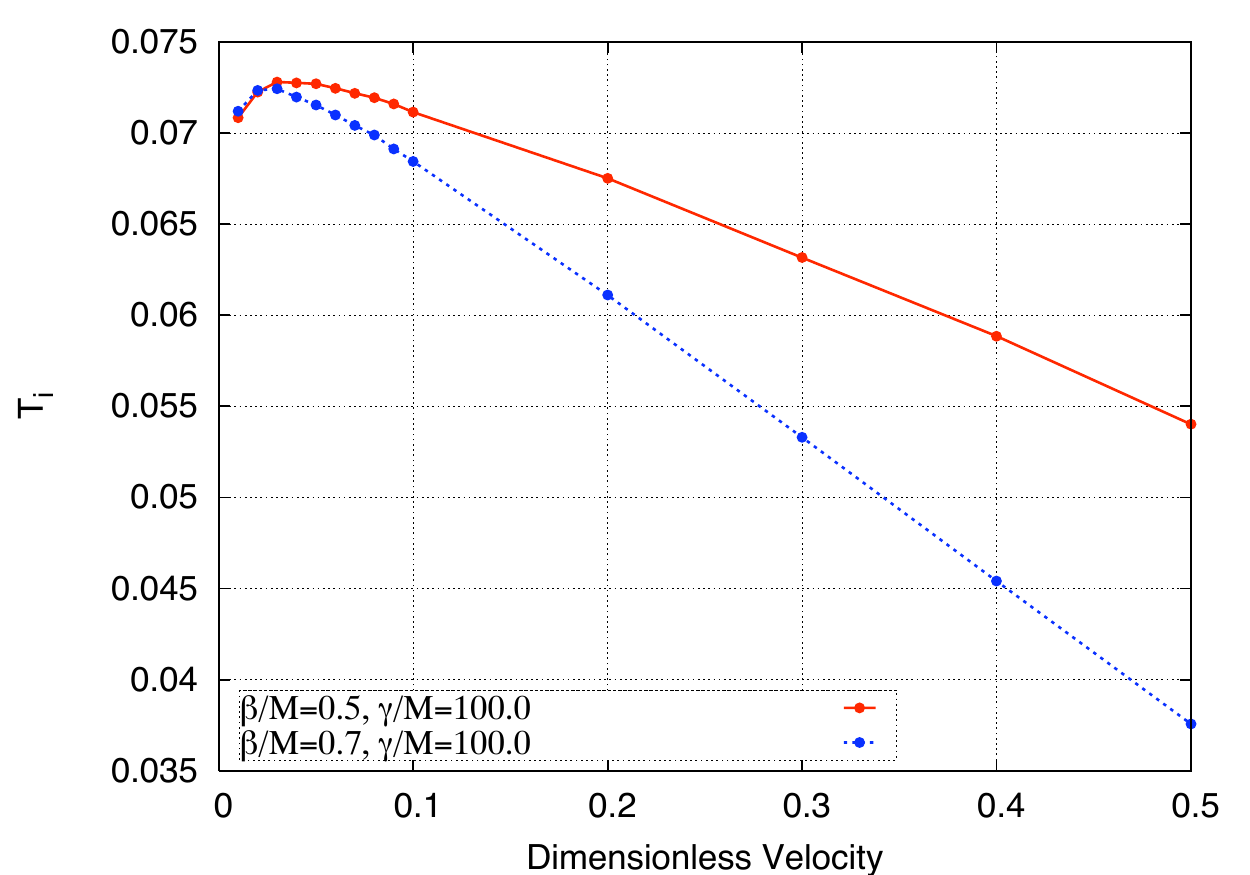}
\caption{The interface temperature vs. the velocity of the interface for two values of $\beta$.}
\label{ti_vel}
\end{figure}
This is shown in Figure~\ref{ti_vel}, which plots the interface temperature versus the pulling speed for two values of $\beta/M$, showing that increasing either speed or $\beta$ has the effect of decreasing the local interface temperature. 

To better understand the role of $\beta$ on interface kinetics, we considered the lowest order sharp interface limit of the amplitude equation corresponding to the density equation used in this study. Details of the derivation are shown in Appendix~\ref{appb}. This approach begins with PFC free 
energy coarse grained in terms of the amplitude of the density field, $\phi$, where $\phi=0$ in the liquid and $\phi=\phi_s$ in the solid, where $\phi_s$ 
is given in Appendix~\ref{appb}. This coarse grained free energy is given by Eq.~\ref{ampfunc}, and was first introduced in \cite{elder2010}.
The equation of motion of the density amplitude, $\phi$, is given by Eq.~\ref{ampdynamics}, for a co-moving reference frame. The addition of the inertial term in 
the dynamics studied here re-scales the coefficient of the highest gradient term, i.e. $C$ in Eq.~\ref{ampdynamics}).  In the limit of small velocities, Appendix~\ref{appb} derives a relation between the interface temperature ($\Delta B_o^i$) versus the steady state interface velocity $V,$ reproduced here for convenience,
\begin{align}
\Delta B_o^i=t\phi_s - B_2^l\psi_l^2-\frac{90}{24}\nu\phi_s^2-\frac{V}{3M\sqrt{C}} \sigma_{\phi},
\label{interfacetemp}
\end{align}
where $C \equiv W(\hat{n})^2-\beta V^2/M$,  $\sigma_{\phi} = \int_{-\infty}^{\infty} \left( \partial_x \phi(x) \right)^2 dx$, $\phi$ is the steady state solution of the phase field equation and $W(\hat{n})$ the anisotropy interface width of $\phi$. 
The precise form of $\phi$ is not known, however we assume assume that it exhibits a transition from $0$ to $1$ across the solid-liquid 
interface, analogous to a hyperbolic tangent  solution, which is the lowest order solution of Eq.~\ref{ampdynamics} in the $V=0$ limit. Numerical solutions of the amplitude equation are consistent with this assumption.  As shown in appendix~\ref{appb} the assumption of a $(1- \tanh(y))/2$ profile leads to $\sigma_\phi=1/3$. The variables  $\psi_l, \psi_s$ are the interface concentrations on the liquid/solid sides of the interface.  
\begin{figure}[htbp]
\resizebox{4.5in}{!}{\includegraphics{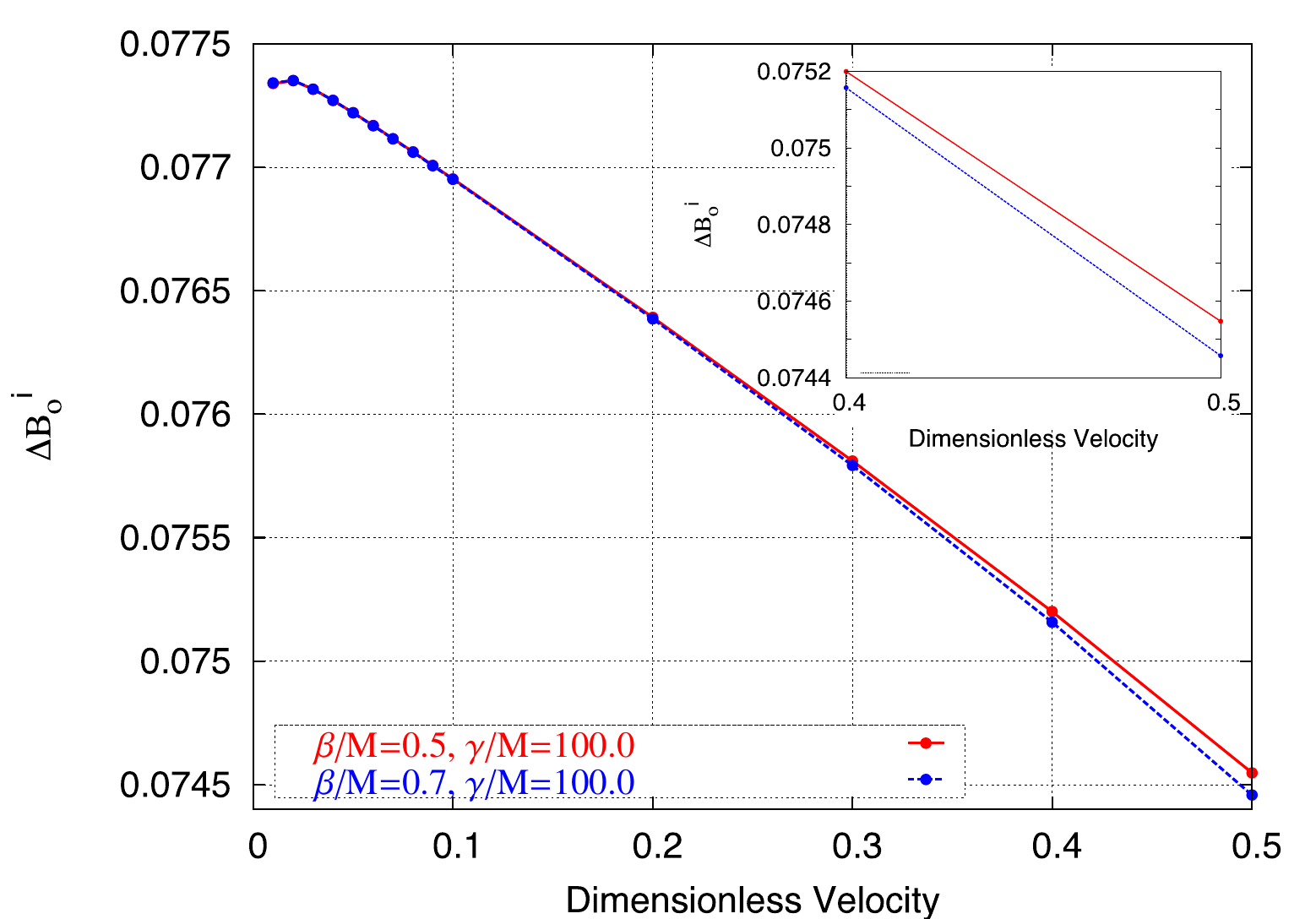}}
\caption{Interface temperature vs. interface velocity using Eq.~\ref{tifinal}. The  $\psi_l$ and $\psi_s$ in Eq.~\ref{interfacetemp} were extracted from the 
numerical simulations of the corresponding $\gamma$ and $\beta$. ($\sigma_{\phi}=1/3$ in Eq.~\ref{interfacetemp}).}
\label{TiFig}
\end{figure}

Figure~\ref{TiFig} plots $\Delta B_o^i$ predicted by Eq.~\ref{interfacetemp} using the parameters from our simulations for two values of $\beta/M$. Eq.~\ref{interfacetemp} has qualitatively similar shape and order of magnitude in the low $V$ range as the data from the numerical simulations. The discrepancy in $\Delta B_o$ for large velocities is a consequence of the crude low-$V$ approximations used to derive Eq.~\ref{interfacetemp}.  Despite the approximate form of Eq.~\ref{interfacetemp}, it allows us to understand the basic role of $\beta$ on interface kinetics. For example, it predicts that increasing $\beta$ at large $V$ leads to a decrease in dimensionless interface temperature. It is noted that while our simulations show that increasing $\beta$ also  increases $\psi_l$ (i.e. decreases 
$k(V)$ in Figure.~\ref{gkv}), the magnitude of $\psi_l^2$ is too small to account for the change in $\Delta B_o^i$ in Figs.~\ref{ti_vel} or \ref{TiFig}  
(Eq.~\ref{interfacetemp}).

Equation~\ref{interfacetemp} is the low-$V$ analogue of the sharp-interface relation derived by Aziz \cite{Azizbot93} for the continuous growth model. The intertial  kinetics of the density equation are seen to essentially renormalize the dimensionless wave speed via $C$ factor in the last term of Eq.~\ref{interfacetemp}. It is noted that at vanishing velocities, $C$ becomes essentially constant independent of $V$. As velocity increases, however, the role of $\beta$ is enhanced. Specifically, Stefanovic et. al showed that $\beta$ increases the effective time the system spends in wave like dynamics. This will have the effect of increasing the effective kinetic term in the sharp interface term in Eq.~\ref{interfacetemp}.
 
It is instructive to derive a more precise form of the interface kinetics for the alloy PFC model that couples information from the equation for $\psi$ with Eq.~\ref{interfacetemp}. This can be done more accurately than what was done here by solving the model equations perturbatively in a small variable like $C$, and matching solution in the interface regions to those outside the interface (i.e. bulk) regions.  This will allow us to simultaneopusly relate the interface temperature $\Delta B_o^i$ to both $\beta$ and $\gamma$. This will be presented elsewhere.

           
\section{Conclusion}
\label{conclusions}
The present study of rapid solidification using a binary PFC model was conducted to understand the role of the inertial dynamics in solute trapping in the PFC alloy formalism. According to Sobolev's prediction~\cite{sob95,sob97}, the hyperbolic form of the concentration diffusion equation predicts complete solute trapping to occur at a finite velocity. Our PFC alloy simulations are consistent with this theory, with the transition to complete trapping being controlled by the ratio $\frac{\gamma}{M}$ (which controls inertial dynamics in the solute field). In particular, Increasing $\gamma$ reduces $V_D^B$, leading to complete trapping at lower velocities. Since our alloy model couples the concentration field to that of the density, the role of rigidity in the density field was shown to modulate the role of inertial dynamics in the solute field.  In particular, increasing the time scale over which the density exhibits inertial dynamics (by increasing the parameter $\beta$ in the density equation), leads to a decrease in the interface temperature, and an increase in concentration on the liquid side of the interface, the latter of which decreases the partition coefficient.   Our numerical results were consistent with a new sharp interface equation we derived to relate the interface temperature to the local interface velocity.
 
  
\appendix
\section{} 
\label{appa}
To compute the values of $K(V)$ in Fig~\ref{gkv} and Fig~\ref{bkv}, we need to calculate the averaged solid and liquid concentrations at the interface. However, in the solid phase, the concentration oscillates at a wave-length equivalent to the inter planar spacing.Therefore, to obtain the values of the averaged concentrations, we employ a Fourier filter to smooth the concentration profiles as shown below:

\begin{equation}
\bar{f_n}=\hat{f_n}e^{(-k^2\epsilon^2)/2}
\label{fourierfilter}
\end{equation}
$n$ is the grid spacing, $\hat{f_n}$ is the fourier transform of the concentration profile, $\bar{f_n}$ is the smoothed fourier transform and $\epsilon$ controls the wave modes of the fourier filter.  We then back transform the smoothed profile to real space  $f_n$. To calculate the minimum error associated with this method, we utilize a similar method used by~\cite{laird98,doral08} where the error was calculated by taking the second order derivative of the smoothed profile and sum all the terms over the length of the simulation box as shown in the following equation:
\begin{equation}
S=\sum_{n=0}^{N}(\delta {f_n})^2
\label{errorlaird}
\end{equation}
where $\delta f_n=(f_{n+1}+f_{n-1}-2f_{n})/dx^2$, which represents the second order derivative of the smoothed concentration profile. As the value of $\epsilon$ increases, the error contributions from $(\delta f_n)^2$ decreases as the profile becomes smoother. 

\begin{figure}[htbp]
\resizebox{5in}{!}{\includegraphics{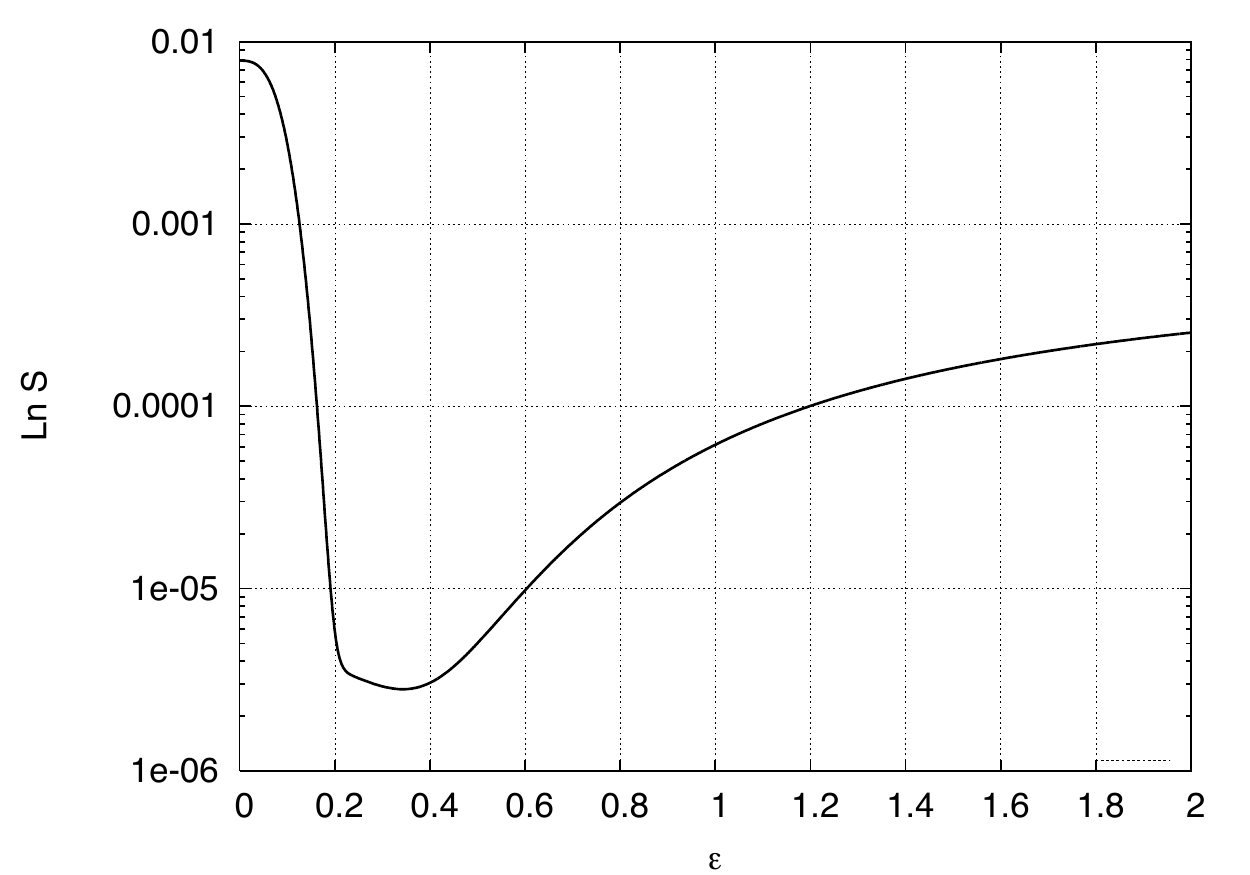}}
\caption{Error dependency on the value of $\epsilon$}
\label{errorfig}
\end{figure}

We have modified Eq.~\ref{errorlaird} by adding error contributions from the interface liquid concentration in the following manner:
\begin{equation}
S=\bigg (\sum_{n=0}^{N}(\delta {f_n})^2 \bigg) + (P_a - P_s)^2
\label{erroreq}
\end{equation}
$P_a$ and $P_s$ are the liquid concentration peaks of the actual concentration profile and the smoothed smoothed profile, respectively. When $\epsilon$ has a small value, the two peaks overlap on top of each, however, as the $\epsilon$ value increases, the smoothed peak deviates from the actual concentration peak causing an increase in the $S$ value. The two terms of Eq.~\ref{erroreq} have error contributions and they vary as we manipulate $\epsilon$.  To minimize the error, we vary the value of $\epsilon$ as illustrated in Fig~\ref{errorfig}. We found an optimal value of $\epsilon=0.343$. 

The systematic error associated with the filter method ($\approx 6\times10^{-6}$) is bigger than the statistical error resulted from the simulations. Since the noise terms were ignored for these simulations, there no is significant statistical error produced. After doing the error propagation analysis the final error associated with $K(V)$ is $\approx1\times10^{-5}$. 

\section{}
\label{appb}
The derivation of the interface temperature begins with the free energy functional of our model, expressed in terms of an amplitude field, $\phi$ and concentration $\psi$. The former is like an order parameter. The form of this free energy  is derived by Elder et al~\cite{elder2010}:
\begin{align}
\freem =\int d^3\vec{x} \bigg\{ W^2(\hat{n}) \bigg(\frac{d\phi}{d u} \bigg)^2+3B_o^x \bigg(\frac{d^2 \phi}{d u^2}\bigg)^2 + 3\Delta B_o \phi^2 - 4t\phi^3+(27-\frac{9}{2})\nu \nline \phi^4 + (\omega + 6B_2^l\phi^2)\frac{\psi^2}{2}+\frac{u}{2}\psi^4+\frac{K}{2}\bigg(\frac{d \psi}{d u}\bigg)^2 \bigg\}
\label{ampfunc}
\end{align} 
where $W^2(\hat{n})$ is the interface width and all the other coefficients are related to PFC model and defined in section~\ref{model}.

The equation of motion for $\phi$ corresponding to the density field in Eq.~\ref{wavedynamics}, written in a steady-state co-moving reference frame, is given by: 
\begin{align}
& C \frac{d^2 \phi}{d \eta^2}-\frac{V}{M}\frac{d \phi}{d \eta} + \pxpy{f}{\phi}=0
\label{ampdynamics}
\end{align}
where $C=W^2(\hat{n})-\frac{\beta V^2}{M}$ and $\pxpy{f}{\phi} = 6[\Delta B_o+B_2^l \psi^2] \phi - 12t\phi^2+90\nu\phi^3$, the latter of which is the derivative of the non-gradient part of the integrand in Eq.~\ref{ampfunc}. To simplify the analysis of Eq.~\ref{ampdynamics} we will consider the low velocity regime, where we can assume to reasonable accuracy that the steady state solution of Eq.~\ref{ampdynamics} can be approximated by:
\begin{align}
&\phi \approx \phi_s g\bigg(\frac{\eta}{\sqrt{C}}\bigg) \nline
&\dxdy{\phi}{\eta} \approx \frac{\phi_s}{\sqrt{C}} \left. \dxdy{g(y)}{y}\right|_{y=\frac{\eta}{\sqrt{C}}}
\label{phi_scaling}
\end{align}
where $\phi_s=(t/15\nu)(1+(\sqrt{1-15(\Delta B_o+B_2^l \psi_s^2)\nu/t^2}))$ is the far field value of $\phi$ in the solid \cite{elder2010}, evaluated here at the capped value of $\Delta B_o$ in the far field solid. The function $g(y)$ satisfies, $g(y \rightarrow \pm \infty)=0,1$. While crude, this approximation allows us to elucidate some important features of the role of $\beta$ in the interface kinetics of our model. 

To extract the relationship between interface temperature and the interface velocity, we employ a projection method,  whereby we multiply both sides of Eq.~\ref{ampdynamics} by $d \phi / d\eta$ and integrate from $-\infty$ to $\infty$, which yields:
\begin{align}
\int_{-\infty}^{\infty} C \dxdy{\phi}{\eta} \dxdy{^2\phi}{\eta^2} d\eta +\frac{V}{M}\int_{-\infty}^{\infty} \bigg(\frac{d \phi}{d \eta}\bigg)^2 d\eta  
-\int_{-\infty}^{\infty}\dxdy{\phi}{\eta}\frac{\partial f}{\partial \phi}d \eta =0
\label{project}
\end{align}
From the boundary conditions of $g(y)$, it is straightforward to show that the first term vanishes.  Using the second of Eq.~\ref{phi_scaling}, the second term becomes 
\begin{align}
\frac{V}{M}\frac{\phi_s^2}{C}\int_{-\infty}^{\infty} \bigg( \dxdy{g(u)}{u}\bigg|_{u=\eta/\sqrt{C}} \bigg )^2 du= \frac{V}{M}\frac{\phi_s^2}{\sqrt{C}} \sigma_\phi
\label{T22}
\end{align}
where $\sigma_\phi \equiv \int_{-\infty}^{\infty} \left( d g / dy \right)^2 dy$. As an example, if we employ $g(y)=\left( 1- \tanh(y) \right)/2$, $\sigma_{\phi}=1/3$. 
Since the last term of Eq.~\ref{project} involves $f(\phi,\Delta B_o(\phi))$, we approximate it by considering the limit where $\sqrt{C} \approx W \ll 1$, 
which allows the lowest order form of amplitude gradient to be approximated by
\begin{equation}
\lim_{W \ll 1} \dxdy{\phi}{\eta} \approx -\phi_s \delta(\eta)
\label{phi_delta}
\end{equation}
This is very approximate but allows the last integral of Eq.~\ref{T22} to be simplified as
\begin{equation}
-\int_{-\infty}^{\infty} \frac{d\phi}{d\eta}\frac{\partial f}{\partial \phi} d\eta \approx \phi_s  \int_{-\infty}^{\infty} \delta(\eta) \frac{\partial f}{\partial \phi} d\eta = \phi_s \frac{\partial f}{\partial \phi}\bigg|_{\eta=0}
\label{T31}
\end{equation}
using the definition above for $\partial{f}/{\partial \phi}$ and evaluating it at the interface, where $\phi(0)=\frac{\phi_s}{2}$ and $\psi(0)=\psi_l$ (interface liquid concentration), results in
\begin{align}
\phi_s \left\{ 6 \left[ \Delta B_o+B_2^l\psi_l^2 \right]  \frac{\phi_s}{2} - 12t \left( \frac{\phi_s}{2} \right)^2+90\nu \left( \frac{\phi_s}{2} \right)^3 \right\}
\label{T32}
\end{align}
Equations~\ref{T22} and \ref{T32} are substituted into Eq.~\ref{project}, yielding
\begin{align}
\Delta B_o^i=t\phi_s - B_2^l\psi_l^2-\frac{90}{24}\nu\phi_s^2-\frac{V}{3M\sqrt{C}} \sigma_{\phi}
\label{tifinal}
\end{align}

\section{Acknowledgment}

The authors like to thank the Natural Science and Engineering Research Council for their financial support of this project. Also, we like to thank Nana Ofori Opoku for useful discussions. 

\pagebreak
\bibliographystyle{plain}


\end{document}